\documentclass{article}

     \PassOptionsToPackage{numbers, compress}{natbib}

     \usepackage[preprint]{neurips_2019}

\usepackage[utf8]{inputenc} 
\usepackage[T1]{fontenc}    
\usepackage{hyperref}       
\usepackage{url}            
\usepackage{booktabs}       
\usepackage{amsfonts}       
\usepackage{nicefrac}       
\usepackage{microtype}      

\usepackage{graphicx}
\usepackage{ amssymb }
\usepackage{amsmath}

\usepackage{xcolor}
\usepackage{caption}
\usepackage{subcaption}

\title{Class-Conditional VAE-GAN for Local-Ancestry Simulation}

\author{%
  Daniel Mas Montserrat\thanks{This work was conducted during an internship at Stanford University.}\\
  Purdue University \\ 
  \And
   Carlos Bustamante \\
  Stanford University \\
  \And
  Alexander Ioannidis \\
  Stanford University \\
}

\begin{document}

\maketitle

\begin{abstract}

Local ancestry inference (LAI) allows identification of the ancestry of all chromosomal segments in admixed individuals, and it is a critical step in the analysis of human genomes with applications from pharmacogenomics and precision medicine to genome-wide association studies. In recent years, many LAI techniques have been developed in both industry \cite{Durand:2014hj} and academic research \cite{maples2013rfmix}. However, these methods require large training data sets of human genomic sequences from the ancestries of interest. Such reference data sets are usually limited, proprietary, protected by privacy restrictions, or otherwise not accessible to the public. Techniques to generate training samples that resemble real haploid sequences from ancestries of interest can be useful tools in such scenarios, since a generalized model can often be shared, but the unique human sample sequences cannot. In this work we present a class-conditional VAE-GAN to generate new human genomic sequences that can be used to train local ancestry inference (LAI) algorithms. We evaluate the quality of our generated data by comparing the performance of a state-of-the-art LAI method when trained with generated versus real data. 

\end{abstract}

\section{Introduction}
Human populations all share a common ancient origin in Africa \cite{DeGiorgio:2009cs}, and a common set of variable sites, but correlations between neighboring sites along the genome, which are typically inherited together, vary between sub-populations around the globe \cite{Li:2008ena}. These correlations along the genome, known as linkage, influence polygenic risk scores (PRS) \cite{Duncan:2019il}, genome-wide association study (GWAS) results \cite{martin2017unexpectedly}, and many other features of precision medicine. Unfortunately, large portions of the world's populations have not been included in modern genetic research studies with over 80\% of these studies to date including only individuals of European ancestry \cite{Popejoy:2016di}. This has serious adverse consequences for the ability of associations learned in these modern studies to be applied to the rest of the world \cite{Duncan:2019il}. Deconvolving the ancestry of admixed individuals using local-ancestry inference can contribute to filling this gap and understanding the genetic architecture and associations of non-European ancestries; thus allowing the benefits of genomic medicine to accrue to a larger portion of the planet's population.

Many methods for local-ancestry inference exist and are open-source, HAPAA \cite{Sundquist2008}, HAPMIX \cite{Price:2009bga} and SABER \cite{Tang2006} infer local-ancestry using Hidden Markov Models (HMMs), LAMP \cite{sankararaman2008estimating} uses probability maximization with a sliding window, and 
RFMix \cite{maples2013rfmix} uses random forests within windows. However, these algorithms all require accessible training data from relevant ancestries in order to recognize those ancestry segments.

The challenge is that many data sets containing human genomic references are proprietary \cite{Han:2017fo, Durand:2015jx}, protected by privacy restrictions \cite{Wojcik:2019hp}, or are otherwise not accessible to the public, especially data sets for under-served or sensitive populations. Generative models that can be easily shared once trained can be useful in such scenarios. While the data sets with their de-anonymizable genome-wide sequences remain securely private, models trained on them could be made publicly available.

In recent years, deep learning has proven effective in solving computer vision and natural language processing problems \cite{nature}. These methods are being used in the biology, medical and genomics fields \cite{Gar2016, lundervold2019overview, li2018heterogeneity, eraslan2019deep}. Specifically, deep learning-based generative methods have been increasingly popular in recent years. Generative networks such as Variational Autoencoders (VAEs) \cite{vae2014} contain a network that encodes the input data into a lower-dimensional space and a decoder that tries to reconstructs the original input. Generative Adversarial Networks (GANs) \cite{goodfellow2014generative} have been able to generate samples that resemble the training data. GANs are able to generate realistic data by using two competing networks: a generator that aims to create realistic new samples and a discriminator that classifies between real and generated samples. Many variants and extensions of GANs and VAEs have been presented recently \cite{arjovsky2017wasserstein, biggan, sohn2015learning}.

\begin{figure}[t!]
        \centering
        \includegraphics[width=0.75\linewidth]{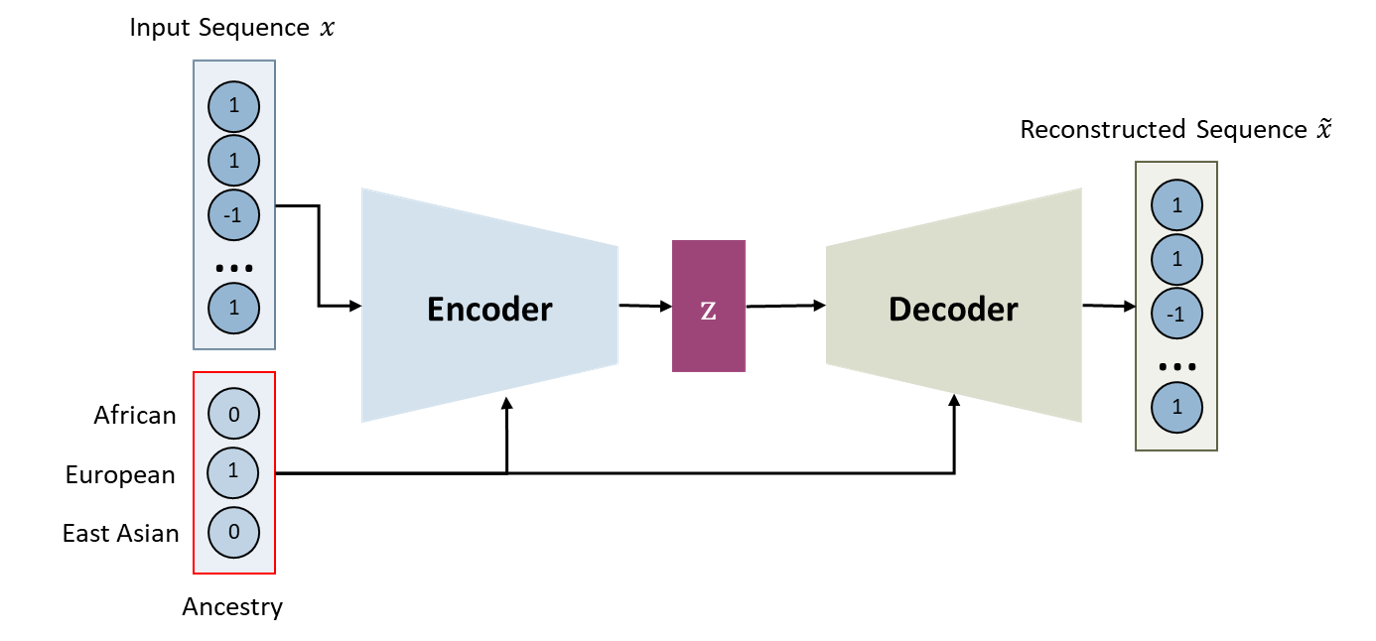}

    \caption{The class-conditional VAE is composed of an encoder-decoder pair. The encoder transforms the input sequence $x$ from the ancestry $c$ into an embedded representation $z$. The decoder transforms the embedding $z$ and ancestry $c$ into a reconstruction of the input sequence, $\Tilde{x}$.}
    \label{fig:vae}
\end{figure}

In this work, we present a class-conditional Variational Autoencoder and Generative Adversarial Network (VAE-GAN) for human genome sequence simulation. The network combines a class-conditional VAE, shown in figure \ref{fig:vae}, with a class-conditional GAN, shown in figure \ref{fig:gan}. The network is able to simulate new single-ancestry sequences that resemble the sequences from the training set. The generated sequences are used to train RFMix.



\section{Out-of-Africa Dataset}

In this work we use simulated datasets with ancestry data generated from out-of-Africa simulations using msprime \cite{kelleher2016efficient}. This simulation models the origin and spread of humans as a single ancestral population that grew instantaneously into the continent of Africa. This population stayed with a constant size to the present day. At some point in the past, a small group of individuals migrated out of Africa and later split in two directions: some founding the present day European populations, and another founding the present day East Asian populations. Both populations grew exponentially after their separation. The parameters that determine the timing of these events, effective population sizes, and growth rates of European and East Asian populations, are presented in Gravel et al. \cite{gravel2011demographic}.

Following the above out-of-Africa model, we generated three groups of 100 diploid individuals of single-ancestry, one group each of African, European and East Asian ancestry. We divided these 300 simulated individuals into training, validation and testing sets with 240, 30 and 30 diploid individuals respectively. Later, the validation and testing individuals were used to generate admixed descendants using Wright-Fisher forward simulation over a series of generations. From 30 single-ancestry individuals, a total of 100 admixed individuals were generated with the admixture event occurring 8 generations in their past to create both validation and testing sets. The 240 single-ancestry individuals were used to train RFMix and the class-conditional VAE-GAN, and the 200 admixed individuals of the validation and testing sets were used to evaluate RFMix following training. Throughout we use chromosome 20 of each individual for experiments.


\section{Network Architecture}




\begin{figure}[t]
        \centering
        \includegraphics[width=0.75\linewidth]{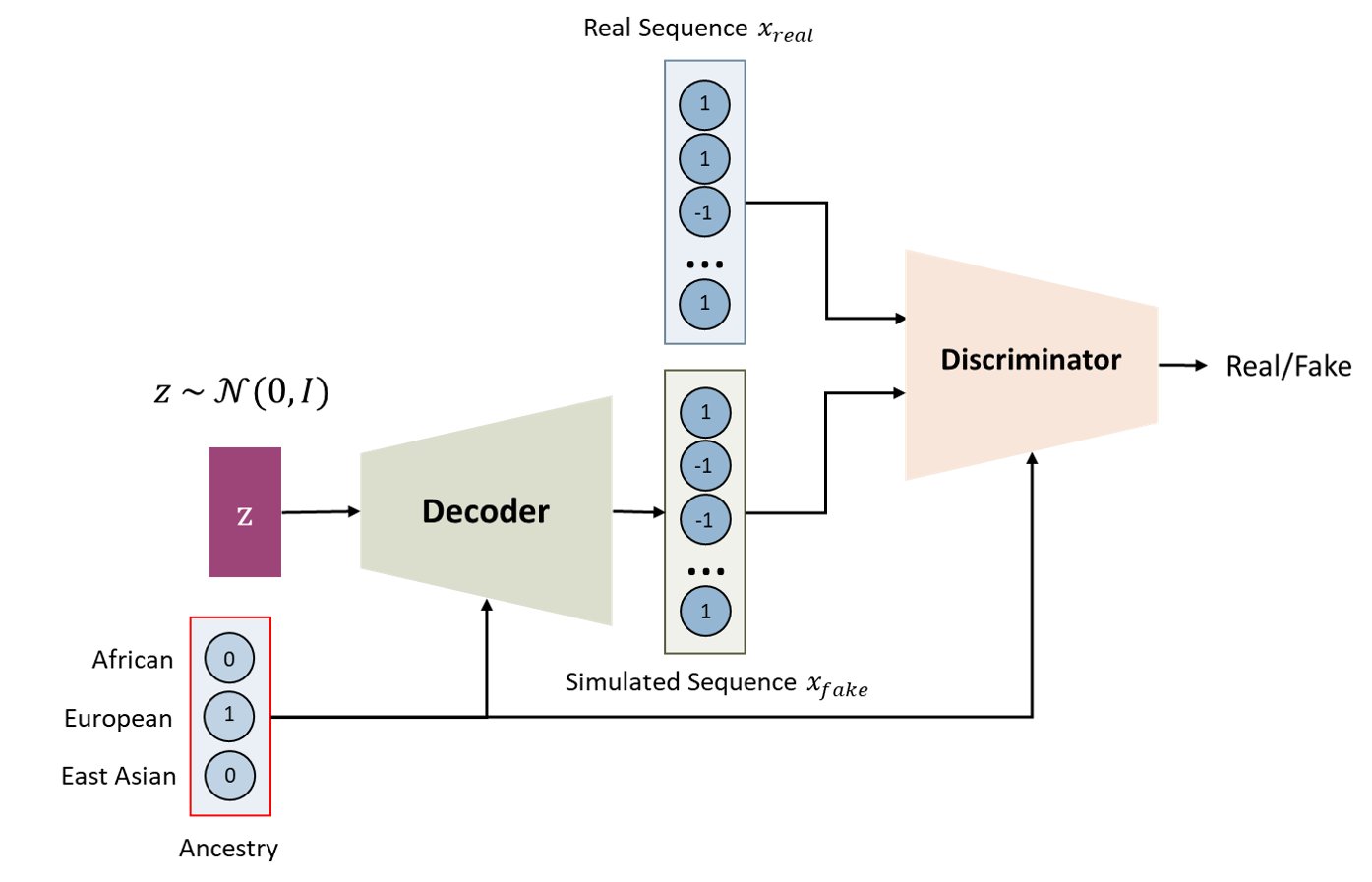}
         \caption{The class-conditional GAN is composed of a decoder-discriminator pair. The decoder generates new samples $x_{fake}$ from a Gaussian representation $z_x$ and ancestry $c$. The discriminator separates between out-of-Africa sequences $x_{real}$ and VAE-GAN generated sequences $x_{fake}$.}
    \label{fig:gan}
\end{figure}

The proposed network splits the genome into fixed-size non-overlapping windows. The SNPs within each window are used as the input for individual class-conditional VAE-GAN's. The input SNPs are encoded as -1 and 1 for each base-pair. Missing input SNPs are modeled by inputting a 0 in the corresponding position. The VAE-GAN's are composed of three sub-networks: an encoder, a decoder, and a discriminator. Each sub-network is class-conditional (i.e. the ancestry is an additional input of the network). The encoder-decoder pair forms a VAE (figure \ref{fig:vae}) while the decoder-discriminator pair forms a GAN (figure \ref{fig:gan}).

The encoder, $q(x; c)$, transforms the input SNPs $x$ from the given the ancestry $c$ (represented with one-hot encoding) into an isotropic Gaussian embedding space $z$. The network encodes the input sequence to the embedding space by estimating $\mu(x;c)$ and $\log\Sigma(x;c)$. The variance is estimated in a logarithmic form to force $\Sigma(x;c)>0$. The embedded representation of a sample $x$ from an ancestry $c$ can be sampled from  $z_x \sim \mathcal{N}(\mu(x;c),\,\Sigma(x;c))$. The sampling can be performed with the reparametrization trick: $z_x = \mu(x;c) + \Sigma(x;c) \odot \epsilon$, where $\epsilon \sim \mathcal{N}(0,\,I)$ and $\odot$ is an element-wise multiplication. The encoder networks begin with an input linear layer of size $(W+C) \times H$, where $W$ is the window's size, $C$ is the number of ancestries, and $H$ is the size of the hidden layer. Following the first layer, a ReLU non-linearity and batch normalization is used. Then, two linear layers are used with dimensions $H \times J$, where $J$ is the dimension of the embedding space, to estimate $\mu(x;c)$ and $\log\Sigma(x;c)$.

The decoder, with a given ancestry $c$ and embedded representation $z_x$, tries to reconstruct the input SNPs $\Tilde{x} = p(z_x;c)$. 
In order to obtain training samples for LAI methods, new sequences can be simulated by selecting the desired ancestry $c$, sampling a random embedding, $z \sim \mathcal{N}(0,\,I)$, and reconstructing the SNP sequence $x_{new} = p(z;c)$.
The decoder networks start with an input layer of size $(J+C) \times H$ followed by a ReLU non-linearity, batch normalization and the output linear layer of size $H \times W$.
The discriminator network is trained to distinguish the real samples from the fake samples $\hat{y} = D(x;c)$.  The discriminator networks start with an input layer of size $(W+C) \times H$ followed by a ReLU non-linearity, batch normalization and the output linear layer of size $H \times 1$.

The encoder is trained by minimizing the mean square error between the input and reconstructed sequences and the Kullback-Leibler divergence. The encoder loss function is as follows:


\begin{equation}
\label{eq:lossq}
  \mathcal{L}_q(x,c) = ||x - \Tilde{x}||_2^2 + \frac{1}{2} \sum_{j}^{J}  \mu_j^2 + \Sigma_j - \log\Sigma_j - 1
\end{equation}

where $x$ and $\Tilde{x}$ are the input and reconstructed sequence respectively, $J$ is the dimension of the embedding space, $\mu_j$ is the $j$th element of $\mu_j(x;c)$ and  $\Sigma_j$ is the $j$th element of the diagonal of $\Sigma_j(x;c)$. The decoder is trained by minimizing the mean square error of the reconstruction and the adversarial loss:

\begin{equation}
\label{eq:lossp}
  \mathcal{L}_p(x,z,c) = ||x - \Tilde{x}||_2^2 + \lambda_1   \log(1-D(p(z;c)))
\end{equation}

where $p(z;c)$ is a simulated sequence from a randomly selected ancestry $c$ and $z \sim \mathcal{N}(0,\,I)$. In our work we select $\lambda_1=0.1$. The discriminator is trained using binary cross-entropy with real, $x$, and simulated data, $p(z;c)$:

\begin{equation}
\label{eq:lossd}
  \mathcal{L}_D(x,z,c) =  - \log(D(x)) - \log(1-D(p(z;c)))
\end{equation}

Because the sequence is generated in a windowed approach, a different ancestry can be assigned to each window, to simulate an admixed individual. However, in this work we focus on single-ancestry individuals. The network is trained to obtain haploid sequences, but by generating pairs of haploid sequences, diploid chromosomes can be simulated. In order to avoid duplicate or very similar individuals, we generate $N$ times the number of desired individuals and compute the pair-wise correlations of the generated sequences. Then, we select the $\frac{1}{N}$ individuals with the lowest average correlation. In this paper we use $N=2$.

\section{Experimental Results}
\label{sec:experimental-results}

We use the single-ancestry out-of-Africa individuals of the training set to train each VAE-GAN. After training the networks, we generate a total of 80 synthetic samples per ancestry and train RFMix. RFMix is then evaluated with the admixed individuals in the validation set. We select the hyper-parameters of the VAE-GAN (window size, hidden layer size and embedding space) and the training parameters (learning rate, batch size and epoch) that provide the highest validation accuracy of RFMix. Finally, we compare the testing accuracy of RFMix when trained with out-of-Africa data versus when trained with data generated with the VAE-GANs. Additionally, we compare the results of including the discriminator and the adversarial loss (VAE-GAN) with only using a VAE.

Table \ref{table:accuracy} shows that RFMix obtains comparable accuracies when trained with out-of-Africa and data simulated data. Accuracy results show that adding the discriminator and the adversarial loss helps the network to learn to simulate human-chromosome sequences that are more similar to the original training data and therefore more useful to train LAI methods, providing a significant increase in accuracy.

\begin{table}[h]
  \caption{Accuracy of RFMix \cite{maples2013rfmix} trained with real and generated data}
  \label{table:accuracy}
  \centering
  \begin{tabular}{l c c}
    \toprule
    \textbf{Method}     & \textbf{RFMix Validation Accuracy} & \textbf{RFMix Testing Accuracy} \\
    \midrule
    \textbf{Out-of-Africa Data} & 97.98\%  & 97.75\%    \\ 
    \textbf{Generated Data (VAE)}     & 93.21\%   & 93.05\%     \\
    \textbf{Generated Data (VAE-GAN)}     & 97.58\%   & 97.72\%     \\

    \bottomrule
  \end{tabular}
\end{table}






\section{Conclusions}
In this work we show a proof of concept for data generation using VAE-GANs. Such networks show promising results with Out-of-Africa simulated data. 
Strong simulation methods allow researchers to work infer ancestry using a wide-range of existing tools without the need for having access to real data from sensitive populations, or from proprietary or protected databases. Besides simulation, generative models have the potential to estimate meaningful representations in the embedding space or to be useful tools for data imputation or reconstruction. 

Future work includes using real humane-genome sequences to train and evaluate our networks and studying how generative models can be used to help interpret the histories of populations. 



\newpage
\bibliographystyle{ieeetr}
\bibliography{mybiblio}

\end{document}